%% file: draft_conf.tex
\RequirePackage{lineno}
\documentclass[aps,prl,preprint,tightenlines,superscriptaddress,showpacs,byrevtex]{revtex4}

\usepackage{graphicx} 
\usepackage{color}
\usepackage{dcolumn}  
\usepackage{tabularx,array}

\newcolumntype{C}{>{\centering\arraybackslash}X}
\usepackage{multirow}
\RequirePackage{hyperref}
\hypersetup{colorlinks=false}

\graphicspath{{ps}}

\begin{document}

\vspace*{3\baselineskip} 

\preprint{\vbox{ \hbox{BELLE-CONF-2001}
}}

\title{ \quad\\[0.5cm]  Measurement of two-particle correlations in hadronic $e^{+}e^{-}$ collisions at Belle}

\input{author-conf2020.tex}

\noaffiliation

\begin{abstract}
\input{Abstract/abstract}

\end{abstract}


\maketitle

\tighten

{\renewcommand{\thefootnote}{\fnsymbol{footnote}}}
\setcounter{footnote}{0}

Two-particle angular correlation functions are a well-established means for the studies of  Quark-Gluon Plasma (QGP) formation in nucleus-nucleus   collisions~\cite{Arsene:2004fa,Back:2004je,Adams:2005dq,Adcox:2004mh}. A ridge-like structure for particle pairs having large differences in pseudorapidity ($\Delta\eta$, where $\eta=-\ln{\tan{\theta/2}}$ with the polar angle $\theta$  defined relative to the counterclockwise beam), but small differences in azimuthal angle ($\Delta\phi$), has been observed in such collisions. This signal is interpreted as a consequence of hydrodynamical expansion of the QGP with fluctuations of the initial density and spatial overlap distributions of the colliding ions~\cite{Ollitrault:1992bk,Alver:2010gr}. The ridge-like signal was also observed in high charged-particle multiplicity events in proton-proton, proton-nucleus, deutron-nucleus and helium-nucleus collisions~\cite{Khachatryan:2010gv,Aad:2015gqa,CMS:2012qk,Abelev:2012ola,Aad:2012gla,Adare:2013piz,Chatrchyan:2012wg,Aaij:2015qcq,Adam:2019woz}. However, the physical origin of the ridge signal in these smaller collision systems is not yet understood~\cite{Dusling:2015gta,Nagle:2018nvi}. In hadron initiated collision systems, the complexity introduced by initial states cannot be easily factored out. A large number of theoretical models based on different underlying mechanisms such as partonic initial state correlations~\cite{Dusling:2013qoz}, final-state interactions~\cite{He:2015hfa,Bierlich:2017vhg} and hydrodynamic medium expansion~\cite{Bozek:2011if} have been proposed to explain the observed ridge-like signal in small systems. 

The measurement of high charged-particle multiplicity events originating from the two-quark system could offer significant insights into the origin of the ridge-like signal~\cite{Nagle:2017sjv}. Recently,  experimental studies have been extended to even smaller collision systems such as electron-ion~\cite{ZEUS:2019jya} and electron-position ($e^+e^-$)~\cite{Badea:2019vey} collisions. No significant ridge-like signal was observed in these measurements. 

Taking advantage of the clean environment in $e^+e^-$ collisions and high-statistics data collected by the Belle detector at KEKB, the analysis is performed for the first time at a center-of-mass energy of $\sqrt{s}=10.52$ GeV, which is $60$ MeV lower than the $\Upsilon(4S)$ resonance, closely following the previous analysis procedure with ALEPH archived data~\cite{Badea:2019vey}. In addition, it is suggested that the measurement of the two-particle correlation function 
can provide new inputs to the phenomenological fragmentation models at the low energy regime.

The Belle detector was a large-solid-angle magnetic spectrometer that consisted of a silicon vertex detector (SVD), a 50-layer central drift chamber (CDC), an array of aerogel threshold Cherenkov counters (ACC), a barrel-like arrangement of time-of-flight scintillation counters (TOF), and an electromagnetic calorimeter (ECL) comprising CsI(Tl) crystals located inside a superconducting solenoid coil that provided a 1.5~T magnetic field.  An iron flux-return located outside of the coil was instrumented to detect $K_L^0$ mesons and muons (KLM).  The detector is described in detail elsewhere~\cite{Abashian:2000cg}.
Overall, a data sample of $31.5~{\rm fb}^{-1}$ is utilized in this analysis.

Hadronic event selections~\cite{PhysRevD.64.072001}, including requirements on event multiplicity and energy sum in the ECL, are adopted to suppress contaminations from two-photon, radiative Bhabha and other QED events.
Particles used in the calculation of the two-particle correlation functions are primary charged tracks, defined in this study as prompt tracks or decay products of intermediate particles with proper lifetime $\tau<1 {\rm~cm}/c$ for the event generator or the transverse component of the decay vertex $V_{r}< 1\rm~cm$ in the event reconstruction. Charged tracks are required to be within the detector acceptance, i.e., $17^\circ\leq\theta\leq150^\circ$, and have a transverse momentum in the center-of-mass frame greater than $0.2$ GeV. 
Impact parameter requirements are adopted to select charged tracks within $\pm 2$ cm with respect to the interaction point in the transverse plane, and $\pm 5$ cm along the $z$ direction.
For pairs of low-momentum curling tracks, the softer one is deemed a duplicated track and is hence removed. Tracks from photon conversion candidates are vetoed, with the latter selected using requirements on the invariant mass and decay-vertex radius of the two tracks.
To eliminate the effects of the nonuniform detection efficiency and misreconstruction bias, efficiency correction factors are derived by the \textsc{EvtGen}~\cite{Ryd:2005zz} and \textsc{pythia}~\cite{2001CoPhC.135..238S} based Belle Monte Carlo (MC) sample, where hadronic $q\bar{q}$ ($q=u,d,s,c$) fragmentation as well as low multiplicity $e^{+}e^{-} \to \tau^{+}\tau^{-}$ and two-photon processes are taken into account.
The MC sample is reweighted to match event multiplicity and thrust spectra of the data in order to correct for the imperfection in MC simulation.
In order to study the multiplicity dependence of the correlation function, the events are binned into multiplicity classes using offline multiplicity, denoted $N_{\rm Trk}^{\rm Offline}$, and counting tracks after all selections.
For low multiplicity events with $N_{\rm Trk}^{\rm Offline}<12$, only a sample size of $11.5~{\rm fb}^{-1}$ is used.
The multiplicity classes used in this study, their corresponding fraction of data, and the mapping of average offline multiplicities $\langle N_{\rm Trk}^{\rm Offline}\rangle$ to average multiplicities after efficiency correction $\langle N_{\rm Trk}^{\rm Corr}\rangle$ are listed in Table~\ref{tab:NtrkCorr}.

\begin{table}[ht]
\begin{center}
\caption{Average multiplicities and corrected multiplicities of different $N_{\rm Trk}^{\rm Offline}$ intervals. 
}
 \begin{tabularx}{0.73\textwidth}{XcX|XcXXcXXcX}
\hline \hline
&$N_{\rm Trk}^{\rm Offline}$ interval &&& Fraction of data (\%) &&& $\langle N_{\rm Trk}^{\rm Offline}\rangle$ &&& $\langle N_{\rm Trk}^{\rm Corr}\rangle$ &\\
\hline 
&$6 \le N_{\rm Trk}^{\rm Offline} < 10$   &&& 43.56 &&& 6.97 &&& 7.05 &\\
&$10 \le N_{\rm Trk}^{\rm Offline} < 12$   &&& 2.58 &&& 10.26 &&& 10.12 &\\
&$12 \le N_{\rm Trk}^{\rm Offline} < 14$   &&& 0.28 &&& 12.20 &&& 11.88 &\\
&$N_{\rm Trk}^{\rm Offline} \ge 12$  &&& 0.29 &&& 12.32 &&& 11.99 &\\
&$N_{\rm Trk}^{\rm Offline} \ge 14$     &&& 0.02 &&& 14.22 &&& 13.72 &\\
\hline \hline
\end{tabularx} 
\label{tab:NtrkCorr}
\end{center}
\end{table}

Following a procedure similar to that has already been established in Ref.~\cite{CMS:2012qk}, the two-particle correlation function observable is defined as
\begin{equation}
\frac{1}{N_{\rm trig}}\frac{d^2N^{\rm pair}}{d\Delta\eta d\Delta\phi}= B(0,0) \times \frac{S(\Delta\eta, \Delta\phi)}{B(\Delta\eta, \Delta\phi)},
\label{eqn:2PCAssociatedYield}
\end{equation}
where $N_{\rm trig}$ denotes the number of trigger particles in the event.
The signal correlation function, $S(\Delta\eta, \Delta\phi)$, counts the yield of particle pairs matched per trigger particle within a single event. The denominator $B(\Delta\eta, \Delta\phi)$ denotes the background correlation distribution, counting the per-trigger-pairing-yield with a ``mixed event''~\cite{Khachatryan:2010gv}. Two functions express the particle pairs' angular separation $\Delta\eta$ and $\Delta\phi$, with the pairing-yields reweighted by efficiency correction factors, and can be explicitly written out as
\begin{equation}
S(\Delta\eta, \Delta\phi) = \frac{1}{N_{\rm trig}}\frac{d^2N^{\rm same}} {d\Delta\eta d\Delta\phi}, \quad 
B(\Delta\eta, \Delta\phi) = \frac{1}{N_{\rm trig}}\frac{d^2N^{\rm mix}}{d\Delta\eta d\Delta\phi},
\label{eqn:2PCSigBkgFunctions}
\end{equation}
where the superscripts ``same'' and ``mix'' indicate that particle pairs are taken from the same and mixed event, respectively.
The mixed event in this work is an average of tracks coming from three random events where $N_{\rm Trk}^{\rm Offline}$ is the same as the event to which the trigger particle belongs.
The $B(0,0)$ factor is incorporated in the calculation of the correlation function, serving as the normalization of the artificially constructed $B(\Delta\eta, \Delta\phi)$. This factor is obtained by extrapolating the function value to the origin of $B(\Delta\eta, \Delta\phi)$.
An additional correction on the correlation function is applied to deal with the effects introduced by using finite-bin histogramming to approximate the density function.
The bin-size effect on the integrated long-range yield is investigated and the magnitude of the correlation function is calibrated.

The two-particle correlation function is explored in two coordinate systems -- beam and thrust axis coordinates in the $e^+e^-$ center-of-mass frame. The former is the same
as those presented in most of the two-particle correlation studies, while in the latter, initiated by Ref.~\cite{Badea:2019vey},  the event thrust axis is used as the reference axis,
with missing momentum of the event included. 
In the viewpoint of relativistic fluid dynamics~\cite{Heinz:2013th}, conventional measurements in beam axis coordinates are sensitive in their transverse directions, probing any anisotropic behaviour among the QCD medium, which are widely studied as the phenomena of elliptic or triangular flow~\cite{Thomas:2010zym, Alver:2010gr,PhysRevLett.105.252302}. In a different kind of search in the $e^{+} e^{-}$ annihilation process, when the interaction medium is located in between or along the color string connecting the $q\bar{q}$, measuring with a coordinate system defined by the event thrust axis provides a more explicable picture. In analogy to the beam axis analysis, correlation functions under thrust axis coordinates are less sensitive to quark-initiated jets close to the thrust axis; however, they offer a better description in the mid-rapidity region, where additional soft gluon emissions apart from the leading quark-antiquark dijet like structure arise in higher order corrections.
In the thrust axis coordinate analysis, the kinematics ($p_{\rm T}, \eta, \phi$) of a mixed event are calculated with respect to the thrust axis of the event in which the trigger particle belongs. For the event mixing in the thrust axis analysis, no further requirements to the mixed events are applied except the multiplicity matching. A reweighting correction for the kinematics distribution of the mixed event is adopted in order to match physical kinematic spectra.

\begin{figure}[ht]
\centering
		\includegraphics[width=0.45\textwidth]{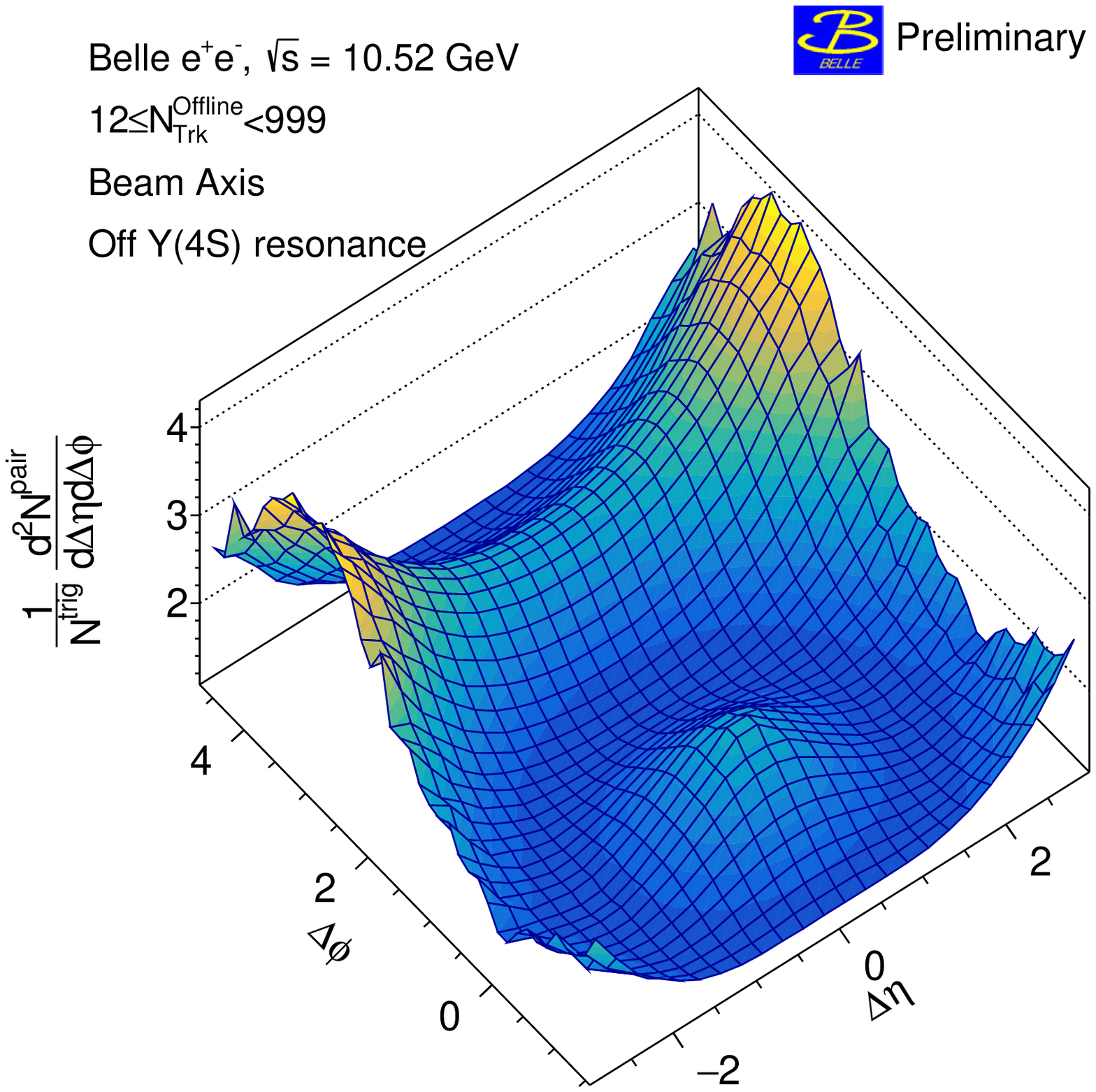}
		\includegraphics[width=0.45\textwidth]{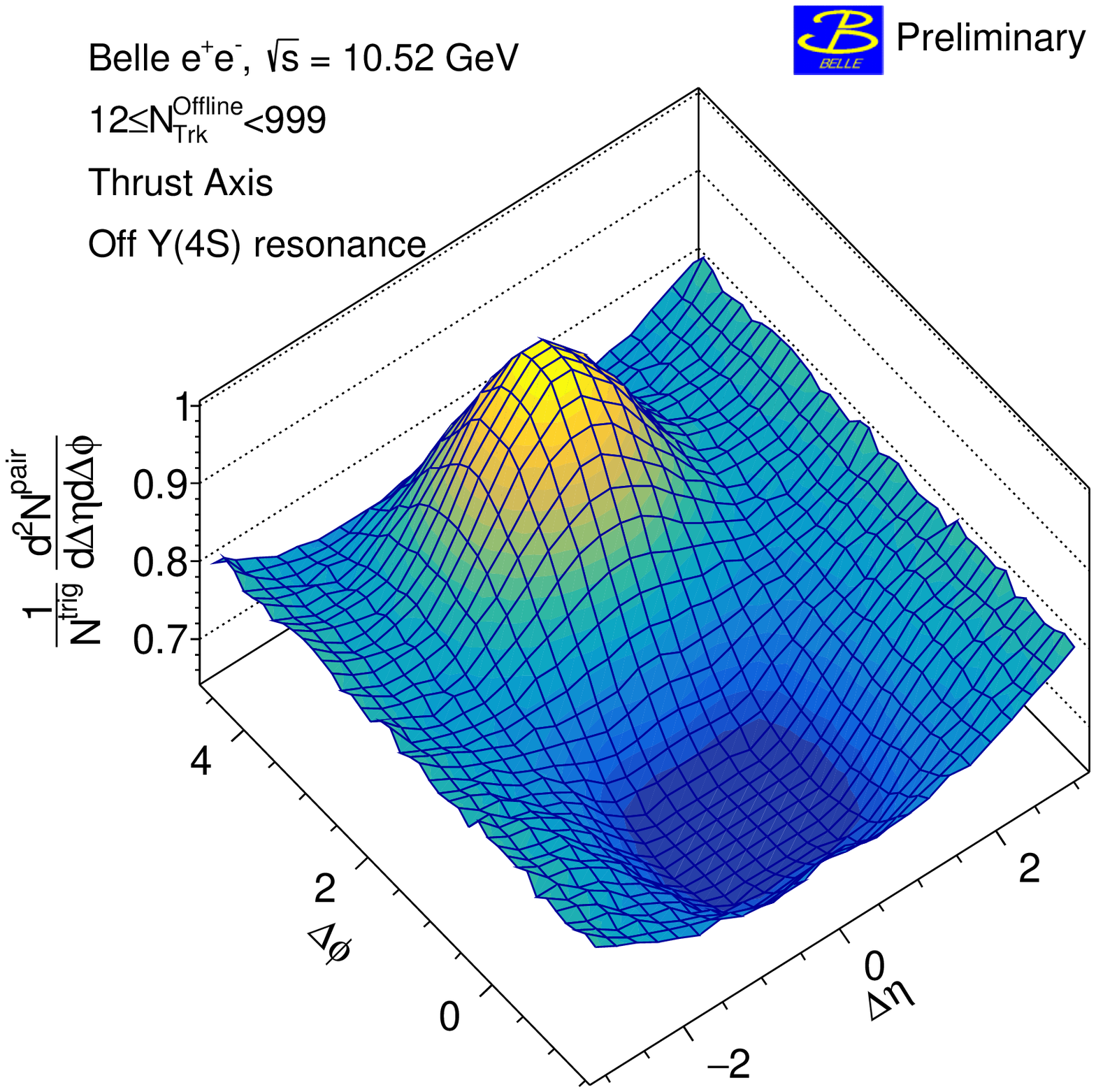}
\caption{Two-particle correlation functions for beam (left) and thrust (right) axis analyses with offline multiplicity $N_{\rm Trk}^{\rm Offline} \ge 12$.}
\label{fig:2PC}
\end{figure}

In Fig.~\ref{fig:2PC}, correlation functions 
with offline multiplicity $N_{\rm Trk}^{\rm Offline} \ge 12$ are shown for both beam and thrust axis coordinates.
In the beam axis coordinate view,
the peak near the origin $(\Delta\eta, \Delta\phi)=(0,0)$ has contributions from pairs originating in the same jet, while the structure at $\Delta\phi\approx\pi$ is from back-to-back correlations. The particles in $e^+e^-$ collisions mainly come from dijet-like $q \bar{q}$ events.
In contrast, for the thrust axis coordinates, the dominant structure is the hill-like bump near $(\Delta\eta, \Delta\phi)\approx(0,\pi)$, while lacking a sizeable near-side correlation.

Evidence for the ridge signal can be best examined in the azimuthal differential yield $Y(\Delta \phi)$ by averaging the correlation function over the long-range region with $1.5 \le |\Delta \eta| < 3.0$.
A ``zero yield at minimum''(ZYAM) method~\cite{Ajitanand:2005jj} is further implemented to separate any enhanced near-side correlation around $\Delta \phi = 0$ as distinct from a constant correlation.
The constant 
contribution along $\Delta \phi$, denoted as $C_{\rm ZYAM}$, is estimated by the minimum of the fit with a third-order Fourier series to the data points.
A fit with a third-order polynomial plus a cosine term, and with a fourth-order polynomial are also checked to estimate $C_{\rm ZYAM}$ in parallel. The discrepancies of different fit evaluations are considered as a source of systematic uncertainties in the final quantification of the ridge signal.

\input{Systematics}

\begin{figure}[ht]
\centering
		\includegraphics[width=0.9\textwidth]{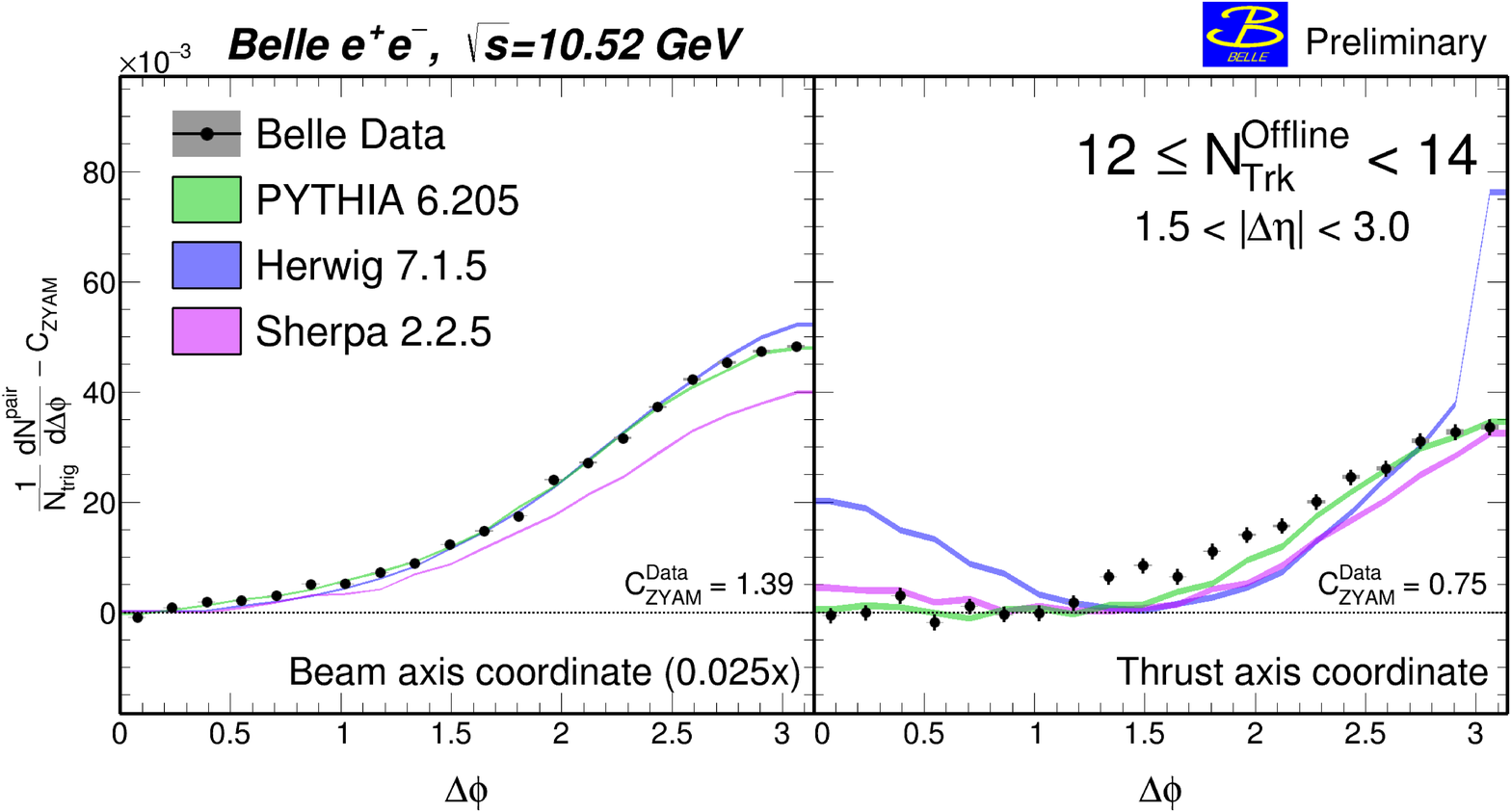}
		\includegraphics[width=0.9\textwidth]{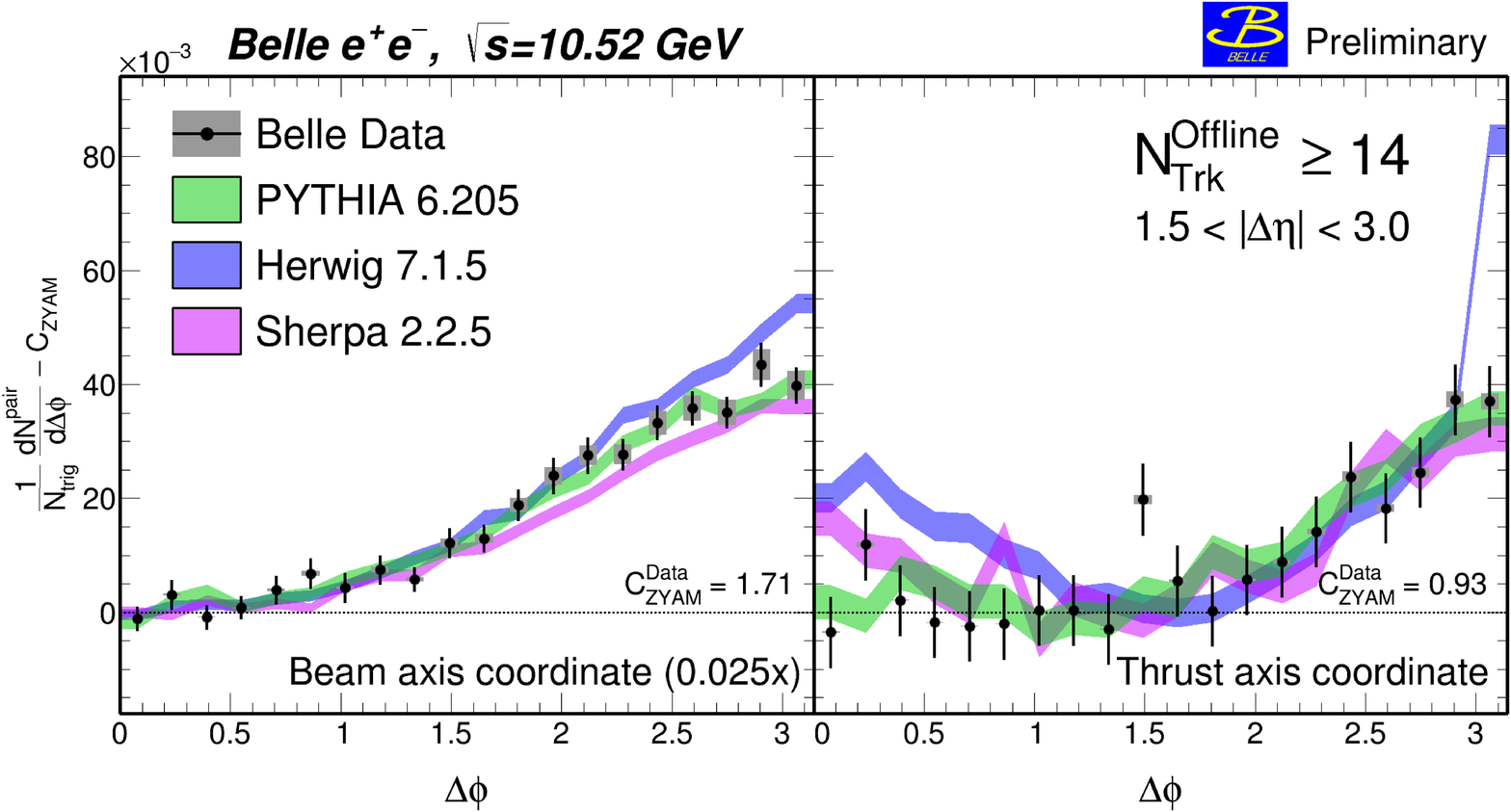}
\caption{Comparison of ZYAM-subtracted $Y(\Delta \phi)$ in the range $1.5 \le |\Delta \eta| < 3.0$ for beam (left) and thrust (right) axis analyses. The colored bands show simulation predictions from Belle MC (green), \textsc{herwig} (blue) and \textsc{sherpa} (violet). The error bars on the data represent the statistical uncertainties, and the gray boxes are systematic uncertainties. }
\label{fig:dNdphi_generator}
\end{figure}

Fig.~\ref{fig:dNdphi_generator} shows the measurement of long-range $Y(\Delta \phi)$ after performing the ZYAM method, along with the comparison of predictions from Belle MC, \textsc{herwig 7.1.5}~\cite{Bellm:2015jjp} and \textsc{sherpa 2.2.5}~\cite{Gleisberg:2008ta} event generators.
The region with small azimuthal angle difference ($\Delta \phi \approx 0 $) is where possible ridge signals would be visible as a nonzero value.
In the beam axis coordinates, all generators are consistent  with data in the near-side ridge region, but deviate in the away-side region.
Under the thrust axis analysis,
the Belle simulation, with specific tunes subjected to Belle data, gives a better description of data. A larger discrepancy from data is seen in the \textsc{herwig} simulation, especially in the near-side ridge-prone region and away-side region.

\begin{figure}[ht]
\centering
    \includegraphics[width=0.5\textwidth]{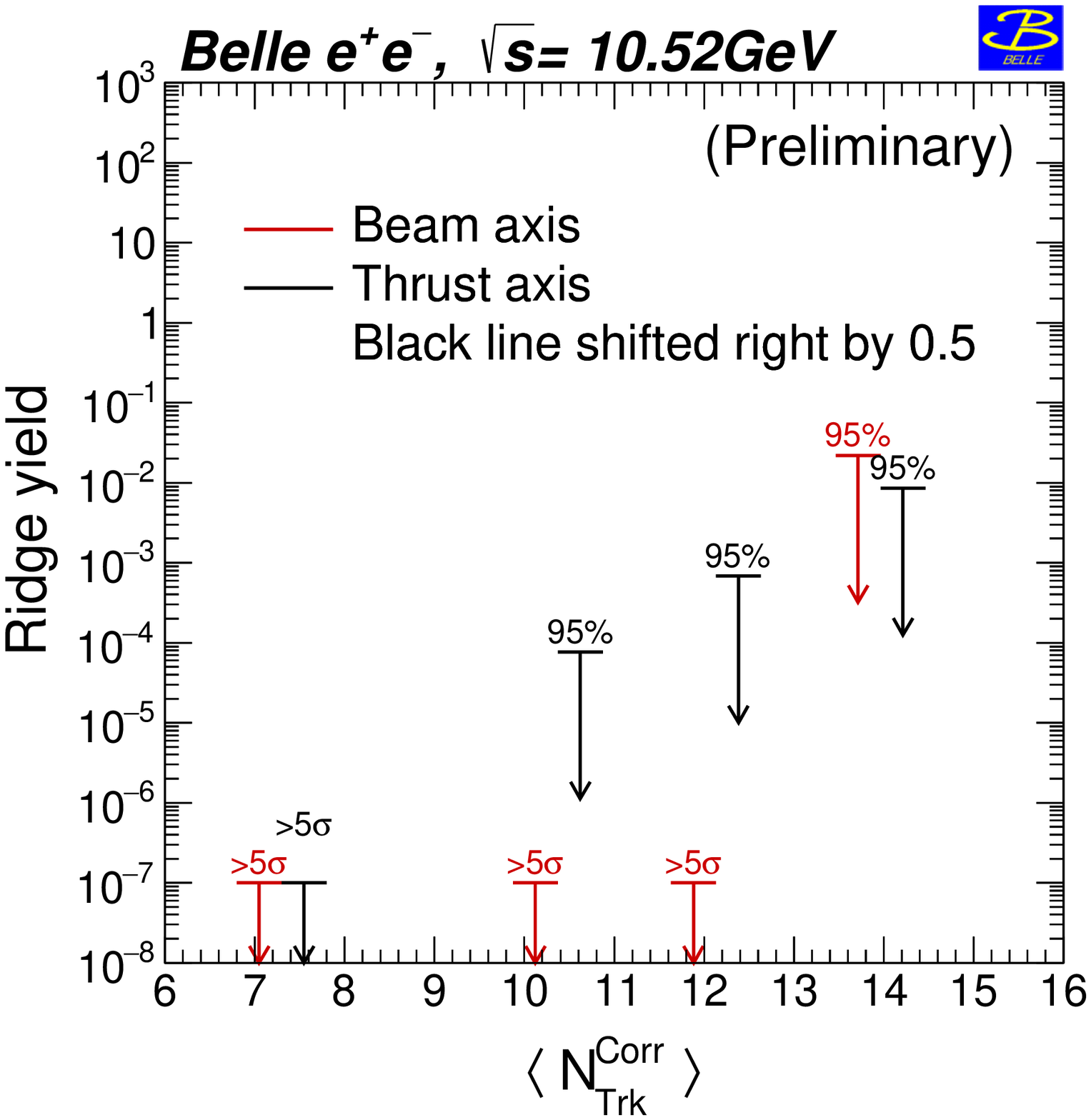}
\caption{Upper limits of the ridge yield as a function of $\langle N_{\rm Trk}^{\rm Corr}\rangle$ in beam axis coordinate (red) and thrust axis coordinate (black) frames. Thrust axis data have been shifted right by 0.5 for presentation purposes. The label ``$>5\sigma$'' indicates the $5 \sigma$ confidence level upper limit.}
\label{fig:CLplots}
\end{figure}

The significance of any ridge signal can be quantified by integrating over $Y(\Delta\phi)$ from $\Delta\phi=0$ to where the ZYAM fit minimum occurs. 
There is no obvious ridge-like structure in either beam or thrust axis analysis. As a result, a bootstrap procedure~\cite{Efron:1979bst} is implemented and the confidence limit of the integrated ridge yield is reported. In the bootstrap procedure, each azimuthal differential yield distribution is varied according to their statistical and systematic uncertainties. Each yield distribution is sampled $2\times10^6$ times, 
and 95\% upper limits are reported. For results with scarce ridge signal, an over $5 \sigma$ significance level for measuring less than $10^{-7}$ ridge yield is quoted, instead.
The upper limits as a function of $\langle N_{\rm Trk}^{\rm Corr}\rangle$ are shown in Fig.~\ref{fig:CLplots}. 

In summary, the first measurement of two-particle correlations using beam and thrust axis coordinate systems, performed using hadronic $e^+e^-$ collisions at $\sqrt{s}= 10.52$ GeV is reported. 
A strong exclusion of ridge yield under the beam axis coordinate frame is set. Compared with measurements in hadronic collision systems and in high energy $e^+e^-$ collisions, the correlation function distribution is heterodox when viewed in the thrust axis coordinate frame, with near-side peak structure, interpreted as intra-jet correlation, not being seen.
Investigations show the magnitude of the near-side peak in the correlation function is strongly correlated with the collision energy when represented under thrust axis coordinates.
95\% upper limits or a $5 \sigma$ exclusion are set for the absence of any ridge yield in our measurement.
Belle MC sample, along with \textsc{herwig} and \textsc{sherpa} event generators are examined.
Similar to the conclusion from the previous analysis with ALEPH archived data, the results in this study are found to be better described by \textsc{pythia} and \textsc{sherpa} than \textsc{herwig}.

\vspace{0.3cm}

We thank the KEKB group for the excellent operation of the
accelerator, the KEK cryogenics group for the efficient
operation of the solenoid, and the KEK computer group and
the National Institute of Informatics for valuable computing
and SINET3 network support. We acknowledge support from
the Ministry of Education, Culture, Sports, Science, and
Technology of Japan and the Japan Society for the Promotion
of Science; the Australian Research Council and the
Australian Department of Education, Science and Training;
the National Natural Science Foundation of China under
contract No.~10575109 and 10775142; the Department of
Science and Technology of India; 
the BK21 program of the Ministry of Education of Korea, 
the CHEP SRC program and Basic Research program 
(grant No.~R01-2005-000-10089-0) of the Korea Science and
Engineering Foundation, and the Pure Basic Research Group 
program of the Korea Research Foundation; 
the Polish State Committee for Scientific Research; 
the Ministry of Education and Science of the Russian
Federation and the Russian Federal Agency for Atomic Energy;
the Slovenian Research Agency;  the Swiss
National Science Foundation; the National Science Council
and the Ministry of Education of Taiwan; and the U.S.\
Department of Energy.

\bibliographystyle{apsrev}
\bibliography{belle2}

\end{document}

%% file: author-conf2020.tex
\noaffiliation
\affiliation{University of the Basque Country UPV/EHU, 48080 Bilbao}
\affiliation{Beihang University, Beijing 100191}
\affiliation{University of Bonn, 53115 Bonn}
\affiliation{Brookhaven National Laboratory, Upton, New York 11973}
\affiliation{Budker Institute of Nuclear Physics SB RAS, Novosibirsk 630090}
\affiliation{Faculty of Mathematics and Physics, Charles University, 121 16 Prague}
\affiliation{Chiba University, Chiba 263-8522}
\affiliation{Chonnam National University, Gwangju 61186}
\affiliation{University of Cincinnati, Cincinnati, Ohio 45221}
\affiliation{Deutsches Elektronen--Synchrotron, 22607 Hamburg}
\affiliation{Duke University, Durham, North Carolina 27708}
\affiliation{University of Florida, Gainesville, Florida 32611}
\affiliation{Department of Physics, Fu Jen Catholic University, Taipei 24205}
\affiliation{Key Laboratory of Nuclear Physics and Ion-beam Application (MOE) and Institute of Modern Physics, Fudan University, Shanghai 200443}
\affiliation{Justus-Liebig-Universit\"at Gie\ss{}en, 35392 Gie\ss{}en}
\affiliation{Gifu University, Gifu 501-1193}
\affiliation{II. Physikalisches Institut, Georg-August-Universit\"at G\"ottingen, 37073 G\"ottingen}
\affiliation{SOKENDAI (The Graduate University for Advanced Studies), Hayama 240-0193}
\affiliation{Gyeongsang National University, Jinju 52828}
\affiliation{Department of Physics and Institute of Natural Sciences, Hanyang University, Seoul 04763}
\affiliation{University of Hawaii, Honolulu, Hawaii 96822}
\affiliation{High Energy Accelerator Research Organization (KEK), Tsukuba 305-0801}
\affiliation{J-PARC Branch, KEK Theory Center, High Energy Accelerator Research Organization (KEK), Tsukuba 305-0801}
\affiliation{Higher School of Economics (HSE), Moscow 101000}
\affiliation{Forschungszentrum J\"{u}lich, 52425 J\"{u}lich}
\affiliation{Hiroshima Institute of Technology, Hiroshima 731-5193}
\affiliation{IKERBASQUE, Basque Foundation for Science, 48013 Bilbao}
\affiliation{University of Illinois at Urbana-Champaign, Urbana, Illinois 61801}
\affiliation{Indian Institute of Science Education and Research Mohali, SAS Nagar, 140306}
\affiliation{Indian Institute of Technology Bhubaneswar, Satya Nagar 751007}
\affiliation{Indian Institute of Technology Guwahati, Assam 781039}
\affiliation{Indian Institute of Technology Hyderabad, Telangana 502285}
\affiliation{Indian Institute of Technology Madras, Chennai 600036}
\affiliation{Indiana University, Bloomington, Indiana 47408}
\affiliation{Institute of High Energy Physics, Chinese Academy of Sciences, Beijing 100049}
\affiliation{Institute of High Energy Physics, Vienna 1050}
\affiliation{Institute for High Energy Physics, Protvino 142281}
\affiliation{Institute of Mathematical Sciences, Chennai 600113}
\affiliation{INFN - Sezione di Napoli, 80126 Napoli}
\affiliation{INFN - Sezione di Torino, 10125 Torino}
\affiliation{Advanced Science Research Center, Japan Atomic Energy Agency, Naka 319-1195}
\affiliation{J. Stefan Institute, 1000 Ljubljana}
\affiliation{Kanagawa University, Yokohama 221-8686}
\affiliation{Institut f\"ur Experimentelle Teilchenphysik, Karlsruher Institut f\"ur Technologie, 76131 Karlsruhe}
\affiliation{Kavli Institute for the Physics and Mathematics of the Universe (WPI), University of Tokyo, Kashiwa 277-8583}
\affiliation{Kennesaw State University, Kennesaw, Georgia 30144}
\affiliation{King Abdulaziz City for Science and Technology, Riyadh 11442}
\affiliation{Department of Physics, Faculty of Science, King Abdulaziz University, Jeddah 21589}
\affiliation{Kitasato University, Sagamihara 252-0373}
\affiliation{Korea Institute of Science and Technology Information, Daejeon 34141}
\affiliation{Korea University, Seoul 02841}
\affiliation{Kyoto Sangyo University, Kyoto 603-8555}
\affiliation{Kyoto University, Kyoto 606-8502}
\affiliation{Kyungpook National University, Daegu 41566}
\affiliation{Universit\'{e} Paris-Saclay, CNRS/IN2P3, IJCLab, 91405 Orsay}
\affiliation{\'Ecole Polytechnique F\'ed\'erale de Lausanne (EPFL), Lausanne 1015}
\affiliation{P.N. Lebedev Physical Institute of the Russian Academy of Sciences, Moscow 119991}
\affiliation{Liaoning Normal University, Dalian 116029}
\affiliation{Faculty of Mathematics and Physics, University of Ljubljana, 1000 Ljubljana}
\affiliation{Ludwig Maximilians University, 80539 Munich}
\affiliation{Luther College, Decorah, Iowa 52101}
\affiliation{Malaviya National Institute of Technology Jaipur, Jaipur 302017}
\affiliation{University of Malaya, 50603 Kuala Lumpur}
\affiliation{University of Maribor, 2000 Maribor}
\affiliation{Massachusetts Institute of Technology, Cambridge, Massachusetts 02139}
\affiliation{Max-Planck-Institut f\"ur Physik, 80805 M\"unchen}
\affiliation{School of Physics, University of Melbourne, Victoria 3010}
\affiliation{University of Mississippi, University, Mississippi 38677}
\affiliation{University of Miyazaki, Miyazaki 889-2192}
\affiliation{Moscow Physical Engineering Institute, Moscow 115409}
\affiliation{Graduate School of Science, Nagoya University, Nagoya 464-8602}
\affiliation{Kobayashi-Maskawa Institute, Nagoya University, Nagoya 464-8602}
\affiliation{Universit\`{a} di Napoli Federico II, 80126 Napoli}
\affiliation{Nara University of Education, Nara 630-8528}
\affiliation{Nara Women's University, Nara 630-8506}
\affiliation{National Central University, Chung-li 32054}
\affiliation{National United University, Miao Li 36003}
\affiliation{Department of Physics, National Taiwan University, Taipei 10617}
\affiliation{H. Niewodniczanski Institute of Nuclear Physics, Krakow 31-342}
\affiliation{Nippon Dental University, Niigata 951-8580}
\affiliation{Niigata University, Niigata 950-2181}
\affiliation{University of Nova Gorica, 5000 Nova Gorica}
\affiliation{Novosibirsk State University, Novosibirsk 630090}
\affiliation{Osaka City University, Osaka 558-8585}
\affiliation{Osaka University, Osaka 565-0871}
\affiliation{Pacific Northwest National Laboratory, Richland, Washington 99352}
\affiliation{Panjab University, Chandigarh 160014}
\affiliation{Peking University, Beijing 100871}
\affiliation{University of Pittsburgh, Pittsburgh, Pennsylvania 15260}
\affiliation{Punjab Agricultural University, Ludhiana 141004}
\affiliation{Research Center for Electron Photon Science, Tohoku University, Sendai 980-8578}
\affiliation{Research Center for Nuclear Physics, Osaka University, Osaka 567-0047}
\affiliation{Meson Science Laboratory, Cluster for Pioneering Research, RIKEN, Saitama 351-0198}
\affiliation{Theoretical Research Division, Nishina Center, RIKEN, Saitama 351-0198}
\affiliation{RIKEN BNL Research Center, Upton, New York 11973}
\affiliation{Saga University, Saga 840-8502}
\affiliation{Department of Modern Physics and State Key Laboratory of Particle Detection and Electronics, University of Science and Technology of China, Hefei 230026}
\affiliation{Seoul National University, Seoul 08826}
\affiliation{Showa Pharmaceutical University, Tokyo 194-8543}
\affiliation{Soochow University, Suzhou 215006}
\affiliation{Soongsil University, Seoul 06978}
\affiliation{University of South Carolina, Columbia, South Carolina 29208}
\affiliation{Stefan Meyer Institute for Subatomic Physics, Vienna 1090}
\affiliation{Sungkyunkwan University, Suwon 16419}
\affiliation{School of Physics, University of Sydney, New South Wales 2006}
\affiliation{Department of Physics, Faculty of Science, University of Tabuk, Tabuk 71451}
\affiliation{Tata Institute of Fundamental Research, Mumbai 400005}
\affiliation{Excellence Cluster Universe, Technische Universit\"at M\"unchen, 85748 Garching}
\affiliation{Department of Physics, Technische Universit\"at M\"unchen, 85748 Garching}
\affiliation{School of Physics and Astronomy, Tel Aviv University, Tel Aviv 69978}
\affiliation{Toho University, Funabashi 274-8510}
\affiliation{Department of Physics, Tohoku University, Sendai 980-8578}
\affiliation{Earthquake Research Institute, University of Tokyo, Tokyo 113-0032}
\affiliation{Department of Physics, University of Tokyo, Tokyo 113-0033}
\affiliation{Tokyo Institute of Technology, Tokyo 152-8550}
\affiliation{Tokyo Metropolitan University, Tokyo 192-0397}
\affiliation{Tokyo University of Agriculture and Technology, Tokyo 184-8588}
\affiliation{Utkal University, Bhubaneswar 751004}
\affiliation{Virginia Polytechnic Institute and State University, Blacksburg, Virginia 24061}
\affiliation{Wayne State University, Detroit, Michigan 48202}
\affiliation{Yamagata University, Yamagata 990-8560}
\affiliation{Yonsei University, Seoul 03722}
  \author{A.~Abdesselam}\affiliation{Department of Physics, Faculty of Science, University of Tabuk, Tabuk 71451} 
  \author{I.~Adachi}\affiliation{High Energy Accelerator Research Organization (KEK), Tsukuba 305-0801}\affiliation{SOKENDAI (The Graduate University for Advanced Studies), Hayama 240-0193} 
  \author{K.~Adamczyk}\affiliation{H. Niewodniczanski Institute of Nuclear Physics, Krakow 31-342} 
  \author{J.~K.~Ahn}\affiliation{Korea University, Seoul 02841} 
  \author{H.~Aihara}\affiliation{Department of Physics, University of Tokyo, Tokyo 113-0033} 
  \author{S.~Al~Said}\affiliation{Department of Physics, Faculty of Science, University of Tabuk, Tabuk 71451}\affiliation{Department of Physics, Faculty of Science, King Abdulaziz University, Jeddah 21589} 
  \author{K.~Arinstein}\affiliation{Budker Institute of Nuclear Physics SB RAS, Novosibirsk 630090}\affiliation{Novosibirsk State University, Novosibirsk 630090} 
  \author{Y.~Arita}\affiliation{Graduate School of Science, Nagoya University, Nagoya 464-8602} 
  \author{D.~M.~Asner}\affiliation{Brookhaven National Laboratory, Upton, New York 11973} 
  \author{H.~Atmacan}\affiliation{University of Cincinnati, Cincinnati, Ohio 45221} 
% ...
\collaboration{The Belle Collaboration}

%% file: Abstract/abstract.tex
The enhancement of charged-particle pairs with large pseudorapidity difference and small azimuthal angle difference, often referred to as the ``ridge signal'', is a phenomenon widely observed in high multiplicity proton-proton, proton-ion and deutron-ion collisions, which is not yet fully understood. 
In heavy-ion collisions, the hydrodynamic expansion of the Quark-Gluon Plasma is one of the possible explanations of the origin of the ridge signal. 
Measurements in the $e^+e^-$ collision system, without the  complexities introduced by hadron structure in the initial state, can serve as a complementary probe to examine the formation of a ridge signal.
The first measurement of two-particle angular correlation functions in high multiplicity $e^+e^-$ collisions at $\sqrt{s}=10.52$ GeV is reported. The hadronic $e^+e^-$ annihilation data collected by the Belle detector at KEKB are used in this study. 
Two-particle angular correlation functions are measured over the full azimuth and large pseudorapidity intervals which are defined by either the electron beam axis or the event thrust as a function of charged particle multiplicity. 
The measurement in the event thrust analysis,
with mostly outgoing quark pairs determining the reference axis, is sensitive to the region of additional soft gluon emissions.
No significant ridge signal is observed with either coordinates analyses. Near side jet correlations appear to be absent in the thrust axis analysis.
The measurements are compared to predictions from various event generators and expected to provide new constraints to the phenomenological models in the low energy regime.

%% file: Systematics.tex
The systematic uncertainties due to the selection and correction operations 
are calculated with respect to  long-range $Y(\Delta \phi)$. 
Hadronic event selection is examined by tightening the energy sum cut in the ECL from $E_{\rm sum}>0.18\sqrt{s}$ to $0.23\sqrt{s}$, yielding a 0--1\% difference we assign as an uncertainty.
The primary particle selection systematic is estimated by making variations in the proper lifetime cut $\tau<1$ cm/$c$ and vertex cut $V_{r}<1$ cm, which gives up to 6.3\% uncertainty.
A 0.35\% uncertainty is quoted for the track reconstruction efficiency.
For high-multiplicity event classes, dominant sources of systematic uncertainties are event selection and primary particle selection, where the estimation of uncertainties suffers from the need of large statistic in order to derive a precise reweighting factor for the efficiency correction and for the mixed events. For low multiplicity classes, the dominant source of uncertainty is due to tracking.
Other uncertainties originate from MC reweighting, the $B(0,0)$ factor, mixed events reweighting, scaling corrections due to bin effect and residual bin effects studied, which are checked to contribute less than 0.5\% uncertainty for each source of uncertainty.